\documentclass[10pt,a4paper]{article}
\usepackage{graphicx}
\usepackage{amsmath}
\usepackage{enumerate}
\usepackage{fullpage}
\usepackage{float} 
\usepackage{courier}
\usepackage[font=small]{caption}
\usepackage[labelfont=bf]{caption}
\usepackage{capt-of}
\usepackage{setspace} 
\usepackage{cite}
\usepackage{hhline}
\begin{document}

\begin{center}
\textbf{\Large Formation of morphogenetic patterns in cellular automata} 
\end{center}
\begin{center}
Manan'Iarivo Rasolonjanahary and Bakhtier Vasiev$^*$
\end{center}
\begin{center}
Department of Mathematical Sciences, University of Liverpool, Liverpool, UK
\end{center} 
\begin{center}
\texttt{$^*$e-mail: bnvasiev@liv.ac.uk} 
\end{center}
\textbf{Abstract.} One of the most important problems in contemporary science, and especially in biology, is to reveal mechanisms of pattern formation.\ On the level of biological tissues, patterns form due to interactions between cells.\ These interactions can be long-range if mediated by diffusive molecules or short-range when associated with cell-to-cell contact sites.\ Mathematical studies of long-range interactions involve models based on differential equations while short-range interactions are modelled using discrete type models.\ In this paper, we use cellular automata (CA) technique to study formation of patterns due to short-range interactions.\ Namely, we use von Neumann cellular automata represented by a finite set of lattices whose states evolve according to transition rules. Lattices can be considered as representing biological cells (which, in the simplest case, can only be in one of the two different states) while the transition rules define changes in their states due to the cell-to-cell contact interactions.\ In this model, we identify rules resulting in the formation of stationary periodic patterns.\ In our analysis, we distinguish rules which do not destroy preset patterns and those which cause pattern formation from random initial conditions.\ Also, we check whether the forming patterns are resistant to noise and analyse the time frame for their formation.\ Transition rules which allow formation of stationary periodic patterns are then discussed in terms of pattern formation in biology.

\section{Introduction}    

Biological pattern formation is a complex phenomenon which is studied experimentally in a number of model organisms~\cite{Wolpert2002} and theoretically by means of various mathematical techniques~\cite{Murray2003, Vasiev2013}.\ One of the model organisms commonly used in biological studies of pattern formation is the fly (Drosophila) embryo~\cite{Hake2003}.\ Patterning in the fly embryo takes place in two directions: along the head-to-tail (antero-posterior) axis where the pattern occurs as a repeated structure formed by segments and along the dorso-ventral axis where patterning results into formation of internal morphological structures.\ It is known that the formation of segments is preconditioned by the formation of spatially periodic patterns of gene expressions which takes place at early stages of embryonic development~\cite{Johnston1992201}.\ There are many other examples of formation of periodic patterns in biology including formation of stripes on skin of animals (zebra) and fish. \par In this work, we focus on formation of periodic patterns which allow discrete representation.\ Discrete models often complement continuous models by giving description of patterning processes on different spatial or time scales.\ For example, the continuous (reaction-diffusion) model of anterio-posterior patterning in fly embryo presented in~\cite{Reka2003} allows to conclude that the modelled units (nuclei) quickly reach their stable equilibrium states and there are only four such states in normally evolving embryo.\ This, in turn, allows to simplify the model by reducing its quantitative details and to describe the considered patterning as evolution of a chain of interacting nuclei on discrete level.\ Another example is given by continuous model of pigmentation patterning on skin of growing reptilian~\cite{Manukyan2017}.\ Due to slow diffusion between skin scales this model also reduces to discrete model for interaction between differently pigmented scales, although the pigmentation of each scale is described by differential equations. Generally, discrete models, as compared to continuous models, often offer a simple explanation for basic properties (such as spatial scaling) of morphogenetic patterns. \par Since we are primarily interested in formation of periodic patterns, we can reformulate the problem and ask the general question: what kind of local interactions can result in formation of a stationary periodic pattern?\ In our study, we use von Neumann's cellular automata (CA) to address this question.\ This model has been invented, more than fifty years ago, by Stanislaw Ulam and John von Neumann~\cite{Vonneumann1948, Vonneumann1966}.\ The model is represented by a collection of cells forming a regular grid which evolves over discrete time steps according to a set of rules based on the state of the cells~\cite{Wainer2010}.\ A detailed description of Neumann's CA can be found in a number of sources~\cite{Yang2010, Wolfram2002}. \par Extensive study of various modifications (involving different numbers of dimensions, allowed states per site and neighbours affecting transitions) of this model have been performed by Wolfram~\cite{Wolfram2002, Wolfram1983, Wolfram1986}.\ He has found that patterns forming in cellular automata fall into one of the following four classes:\ 1.\ Homogenous, 2.\ Periodic structures, 3.\ Chaotic structures and 4.\ Complex.\ Correspondingly, all automata also fall into four classes, although the class in which the given automata falls depends on the imposed initial conditions. \par In this work, we will explore rules resulting in the formation of periodic patterns in the CA where each cell can only be in two distinct states.\ This is an extension of the research reported in~\cite{Wolfram2002, Wolfram1983} in two ways: firstly, we consider all 256 rules (while the classification made by Wolfram was restricted by 32 so called ``legal" rules) and, secondly, we are interested in periodic patterns forming in a domain of finite size, and therefore the found rules appear to fall into classes 2 and 4 identified by Wolfram.\ Although this model is very simple, the results obtained allow conclusions to be drawn on the mechanisms of periodic pattern formation in biology, for example the anterio-posterior patterning in fly embryo.

\section{Chain of logical elements}

The simplest version of von Neumann's cellular automata was designed by Wolfram~\cite{Wolfram2002, Wolfram1983} and called ``elementary cellular automaton" (ECA).\ It consists of a regular lattice of cells which form a chain and can only be in two states.\ The position of a cell in the chain will be denoted by the symbol $i$, the total number of cells - by $n$ and the state of cell by $s_i$ ($s_i=0$ or $s_i=1$).\ The state $s_i^{t+1}$ of a cell $i$ at time $t+1$ is determined by the states of cells $i-1$, $i$ and $i+1$ at time $t$ according to a transition rule, defined as $s_i^{t+1}=\left(s_{i-1}^t, s_i^t,s_{i+1}^t\right)$. A cell interacting with its two closest neighbours leads to consideration of $2^3=8$ possible configurations, which are 000, 001, 010, 011, 100, 101, 110 and 111.\ For each of these 8 configurations, the resulting states can be represented as:\ 0$\boldsymbol{s_0}$0, 0$\boldsymbol{s_1}$1, 0$\boldsymbol{s_2}$0, 0$\boldsymbol{s_3}$1, 1$\boldsymbol{s_4}$0, 1$\boldsymbol{s_5}$1, 1$\boldsymbol{s_6}$0 and 1$\boldsymbol{s_7}$1.\ The complete set of middle elements of all resulting states forms a binary number $\boldsymbol{s_7s_6s_5s_4s_3s_2s_1s_0}$ which varies between 0 and 255. In other words, the configurations above allow $2^8=256$ possible rules or cellular automata. The binary number, translated into decimal, is taken to be the rule number. \par Different rules applied to different initial conditions can result into formation of various nontrivial spatio-temporal patterns including oscillations and propagating waves.\ Among these 256 rules, a few symmetry groups can be distinguished so that the rules belonging to one group result in the formation of similar patterns.\ These groups contain up to four rules and their symmetry-based relationships termed complement, mirror image and mirror complement~\cite{Wolfram1986}.\ The definitions of these relationships are based on certain features of the binary representation of rules, namely:\par The complement of a given rule is a rule whose binary representation is obtained from the binary representation of the given rule where (a) the digits are taken in the reverse order and (b) each digit is interchanged with the opposite digit. \par The mirror image of a given rule is the transition rule such that if the given rule contains an elementary rule $f(b, c, d)=a$ then the mirror image contains an elementary rule $f(d, c, b)=a$, that is, the same output is allocated to the rule with mirrored states of neighbours.\par The mirror complement is a complement to a mirror image.\par Our study is performed using the one-dimensional CA described above.\ Note that this model can also be seen as a chain of logical elements.\ It is represented by a onedimensional regular lattice of elements which can be in one of two states.\ The state $S_i$ of the element $i$, can be either ``false" (commonly denoted by ``0" and illustrated in black colour) or ``true" (denoted by ``1" and drawn in white). The state of each element changes with each time increment (except for the border cells which remain unaltered) according to the applied rules. \par The goal of this study is to find out what rules can cause the formation of stationary periodic patterns in a chain of finite size where the states of border elements are fixed.\ We also check whether the periodic patterns are stable,\ i.e.\ noise sensitive.\ To do so, we introduce a noise by changing the state of each element randomly with a certain (small) probability.\ In application to biology, each logical element corresponds to a biological cell and the cell's state reflects the expression of certain genes.\ The ECA represents a two-state model and can only be used for study of patterns formed by expression of a single gene along cells forming a chain.

\section{Results} 

Our study of periodic patterns forming in the ECA has been performed in a few steps.\ First, we have focused on two-periodic patterns and identified the rules which (1) do not destroy a preset two-periodic pattern, (2) allow recovery of two-periodic patterns perturbed by noise and (3) allow formation of two-periodic patterns from random initial conditions.\ Then, we have repeated the above study for the case of three- and more- periodic patterns.

\subsection{Conservation of preset two-periodic patterns in ECA}  

\begin{figure}[H]
\centering
\includegraphics[width=1.0\textwidth]{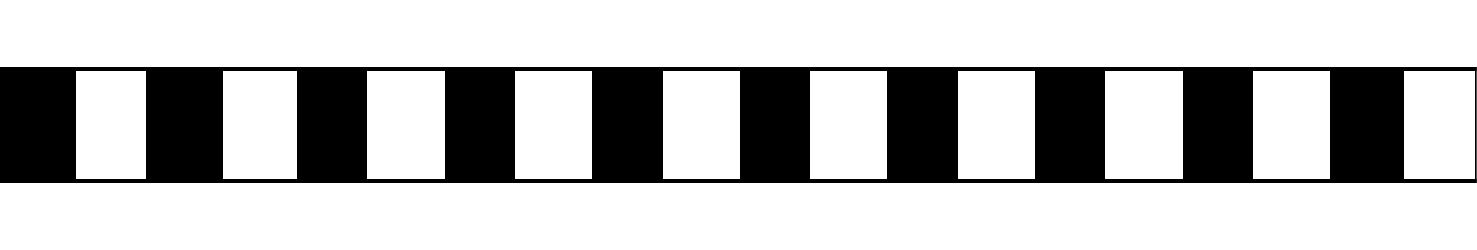}
\caption{Two-periodic pattern in the ECA composed by 20 cells.\ The binary expression for all these rules has the form: $xx\boldsymbol{0}xx\boldsymbol{1}xx$, where $x$ can be either 0 or 1.\ That is, the pattern consisting of a repetition of black and white remains the same when the transition rule leaves the configurations 3 and 6 unaltered:\ $(010) \to (0\boldsymbol{1}0)$ and $(101) \to (1\boldsymbol{0}1)$.}
\label{fig1}
\end{figure}

Our first set of simulations aimed to find all transition rules which do not destroy pre-existing periodic patterns.\ These simulations have been started with preset two-periodic patterns (1 white + 1 black) like the one shown in Fig~\ref{fig1} but with more (usually 60 or more) elements.\ Starting with two-periodic initial conditions, we have found that 64 rules out of 256 (25\%) do not destroy it.\ Their values, in decimal form, are as follows:
\begin{center}
4, 5, 6, 7, 12, 13, 14, 15, 20, 21, 22, 23, 28, 29, 30, 31, 68, 69, 70, 71, 76, 77, 78, 79, 84, 85, 86, 87, 92, 93, 94, 95, 132, 133, 134, 135, 140, 141, 142, 143, 148, 149, 150, 151, 156, 157, 158, 159, 196, 197, 198, 199, 204, 205, 206, 207, 212, 213, 214, 215, 220, 221, 222 and 223.
\end{center} 

The second set of simulations was aimed at finding all rules which are such that, starting with periodic initial conditions (see Fig~\ref{fig1}) and applying a noise, the perturbed pattern is able to recover. The noise level of 0.1\% has been applied meaning that the state of each cell can be altered with probability of 0.001 (i.e.\ one out of 1000 cells is altered each time step).\ Simulations show that 33 out of the above 64 rules (52\%) would allow resistance to the perturbation caused by the noise. These rules are given by the following numbers:
\begin{center}
6, 13, 14, 15, 20, 28, 30, 69, 70, 77, 78, 79, 84, 85, 86, 92, 93, 134, 135, 141, 143, 148, 149, 156, 157, 158, 159, 197, 198, 199, 213, 214 and 215.
\end{center} 
\par The recovery of a perturbed pattern can happen in two ways:\ locally, when the perturbation disappears without affecting surrounding cells, or globally, when the perturbation propagates along the medium to one of its borders and then vanishes.\ Also, for both scenarios there are exceptional rules (28, 70, 78, 92, 141, 156, 157, 197, 198 and 199) such that if the perturbed cell is located next to the border cell then it cannot recover (unless it is hit by another perturbation).

\subsection{Local recovery of perturbed two-periodic patterns}
  
The following rules allow the local extinction of perturbations:
\begin{center}
13, 28, 69, 70, 77, 78, 79, 92, 93, 141, 156, 157, 197, 198 and 199.
\end{center} 
\par For the analysis of the recovery processes, we point out that if the probability of perturbation is small, then the probability that two neighbouring cells are perturbed simultaneously is much smaller and, therefore, can be neglected. With this in mind, let us consider the recovery of the periodic pattern in a chain described, for example, by the rule 77, whose binary number is 01001101. Consider the periodic sequence $\ldots010\boldsymbol{1}01\ldots$ and the scenario when, for example, the element in the 4\textsuperscript{th} position changes to \textbf{0} due to the perturbation caused by noise.\ Then, the above sequence will become $\ldots 010\boldsymbol{0}01\ldots$ .\ We want to know how this sequence will evolve over time and how many time steps it will take for the pattern to recover.\ The changed element can only affect its two closest neighbours, i.e.\ the zeros in 3\textsuperscript{rd} and 5\textsuperscript{th} positions.\ The 3\textsuperscript{rd} zero is in the middle of 10\textbf{0} and will not change following the transition $(10\textbf{0}) \to (10\textbf{0})$. The perturbed cell is in the middle of 0\textbf{0}0 and therefore will change to \textbf{1} following the transition $(0\textbf{0}0) \to (0\textbf{1}0)$.\ The 5\textsuperscript{th} zero is in the middle of \textbf{0}01 and will remain unaltered following the transition $(\textbf{0}01) \to (\textbf{0}01)$.\ Thus, we conclude that it takes one time step for the pattern to restore. The above analysis is valid for all perturbations changing any ``1" to ``0"  inside the periodic chain and therefore it will take one time step for the pattern to recover from all these perturbations. Similarly, one can show that perturbations changing any ``0" to ``1" inside the chain will also vanish in one time step. From here, we conclude that under rule 77 it takes one time step for two-periodic pattern to recover from any single perturbation. Since rule 77 coincides with its own complement, mirror image and mirror complement (rule 77 is unique in this respect), the above result cannot be extended to any other rule. \par There are rules which allow the local recovery of perturbed patterns in more than one time step. Let us, for example, consider the rule 13 (00001101 in binary representation) applied to the preset periodic pattern (i.e.\ $\ldots0101010\ldots$) in a medium of arbitrary (finite) size. It is easy to show that if the noise changes any of the 1s to 0 in the periodic sequence it will take only one time step to recover. On the other hand, it takes two time steps to restore when the noise changes any of 0s to 1.\ To illustrate this, let us consider the finite periodic sequence $\ldots01010101\ldots$ . where the noise changes the digit 0 at the 3\textsuperscript{rd} position into 1, so that the sequence becomes $\ldots01110101\ldots$ . The 1 in the second position will stay unaltered following the transition: $(011) \to (011)$.\ The 1 in the third position will change to 0 following the transition $(111)\to (101)$. The 1 in the fourth position will change to 0 following the transition $(110) \to (100)$. So, after the first step we get the sequence $\ldots01000101\ldots$ . This is identical to the case (considered above) when the state of the $\left(4\textsuperscript{th}\right)$ cell in periodic pattern is changed from ``1" to ``0".\ Its recovery takes only one time step and thus it takes two time steps for the pattern to recover when, owing to the noise, any of 0s change to 1.\ The stated result is exactly the same for the rule 69, which is the mirror image of the rule 13, and it is slightly amended for the rules 79 and 93, which are complements of rules 13 and 69 respectively.\ Namely, in the cases of rules 79 and 93 it takes one time step to recover when any ``0" is changed to ``1" in the preset periodic pattern and two time steps when any ``1" is changed to ``0".

\begin{center}
\begin{tabular}{|c|c|c|c|} 
\hline
\textbf{Rule} & \textbf{Complement} & \textbf{Mirror image} & \textbf{Mirror complement} \\
\hline
\multicolumn{4}{|c|}{\textbf{77}}  \\
\multicolumn{4}{|c|}{$1 \to 0$ (1 TS)} \\
\multicolumn{4}{|c|}{$0 \to 1$ (1 TS)} \\
\hline
\textbf{13} & \textbf{79} & \textbf{69} & \textbf{93} \\ 
$1 \to 0$ (1 TS) & $1 \to 0$ (2 TS) & $1 \to 0$ (1 TS) & $1 \to 0$ (2 TS) \\ 
$0 \to 1$ (2 TS) & $0 \to 1$ (1 TS) & $0 \to 1$ (2 TS) & $0 \to 1$ (1 TS) \\ 
\hline
\textbf{70} & \textbf{157} & \textbf{28} & \textbf{199} \\ 
$1 \to 0$ (2 TS) & $1 \to 0$ (3 TS) & $1 \to 0$ (2 TS) & $1 \to 0$ (3 TS) \\ 
$0 \to 1$ (3 TS) & $0 \to 1$ (2 TS) & $0 \to 1$ (3 TS) & $0 \to 1$ (2 TS) \\ 
(\ldots000) & (\ldots111) & (000\ldots) & (111\ldots) \\
\hline
\textbf{78} & \textbf{141} & \textbf{92} & \textbf{197} \\ 
$1 \to 0$ (3 TS) & $1 \to 0$ (1 TS) & $1 \to 0$ (3 TS) & $1 \to 0$ (1 TS) \\ 
$0 \to 1$ (1 TS) & $0 \to 1$ (3 TS) & $0 \to 1$ (1 TS) & $0 \to 1$ (3 TS) \\ 
(\ldots000) & (\ldots111) & (000\ldots) & (111\ldots) \\
\hline
\textbf{156} & \multicolumn{2}{|c|}{\textbf{198}} & \textbf{156} \\ 
$1 \to 0$ (2 TS) & \multicolumn{2}{|c|}{$1 \to 0$ (2 TS)} & Same as in the first column. \\ 
$0 \to 1$ (2 TS) & \multicolumn{2}{|c|}{$0 \to 1$ (2 TS)} & \\ 
(\ldots111)(000\ldots) & \multicolumn{2}{|c|}{(\ldots000)(111\ldots)} & \\ 
\hline
\end{tabular}
\captionof{table}{Summary of recovery dynamics for rules allowing local recovery of perturbed two-periodic pattern.\ The notation ``$a \to b$ ($n$ TS)" is used to state that the perturbed by noise cell recovers from state ``$a$" to its original state ``$b$" in $n$ time steps.\ The exceptional cases, when the recovery doesn't take place, (and when the noise hits the cell next to the border cell) are described in the extra (last) line. For example, the notation ``(\ldots000)" indicates that the rule doesn't allow recovery in the case when the perturbed chain ends with three 0s.}
\label{table1}
\end{center}

The process of recovery of a perturbed periodic pattern takes even longer for some other rules. For example, in the case of rule 70 whose binary representation is 01000100, it takes two time steps for restoration when the noise changes (almost) any ``1" to ``0" in the preset periodic pattern and three time steps after changing any ``0" to ``1".\ There is an exceptional case for this rule when the perturbed pattern does not recover.\ Namely if, in the medium with a preset periodic pattern such that the states of the rightmost cells are given as (010), the noise strikes the second cell from the right border (resulting into \ldots101000) then the pattern does not recover.\ This is because rule 70 involves the transition $(100) \to (100)$, which allows for the 3\textsuperscript{rd} cell from the right border remaining unaltered, and $(000) \to (000)$, for which the state of the second (perturbed) cell remains unaltered.\ The only way for the periodic pattern to recover is that the noise strikes the same cell once again.\ The results for rule 70 can, with small modifications, be extended to rules 157, 28 and 199 which are respectively complement, mirror image and mirror complement of rule 70.\ The properties of all rules which allow local recovery of perturbations on preset two-periodic patterns are given in Table~\ref{table1}.

The rules listed in Table 1 allow local recovery from any noise, namely from 0 to 1, as well as from 1 to 0 perturbations. However, there are rules which allow the recovery of periodic patterns only from one type of perturbation but NOT from the other and which have not been included in Table 1. These are rules 5, 76, 94, 95, 133 and 205. Rules 76, 94 and 95 allow the recovery of periodic pattern when, due to the noise, 0 has changed to 1. Rules 5, 133 and 205 allow the recovery in case when, due to the noise, 1 has changed to 0. All six rules, when they allow the recovery of perturbed pattern, do it in only 1 time step. 

\subsection{Recovery of two-periodic patterns by means of propagating waves}

In this section, we consider a set of rules which also allow the recovery of perturbed periodic patterns but in a different manner: namely, the perturbation is driven to one of the two edges of the chain where it commonly disappears.\ In the most common scenario, the perturbation moves like a wave with a constant speed and the time required for the perturbation to get to the chain edge is proportional to the initial distance from the perturbed cell to the corresponding edge of the chain.\ However, for some rules the waves of perturbation which form are not regular, so that their speed and the size of perturbed area change over time.\ The rules exhibiting propagating waves of perturbation are:
\begin{center}
6, 14, 15, 20, 30, 84, 85, 86, 134, 135, 143, 148, 149, 158, 159, 213, 214 and 215.
\end{center}

\begin{figure}[H]
\centering
\includegraphics[width=0.72\textwidth]{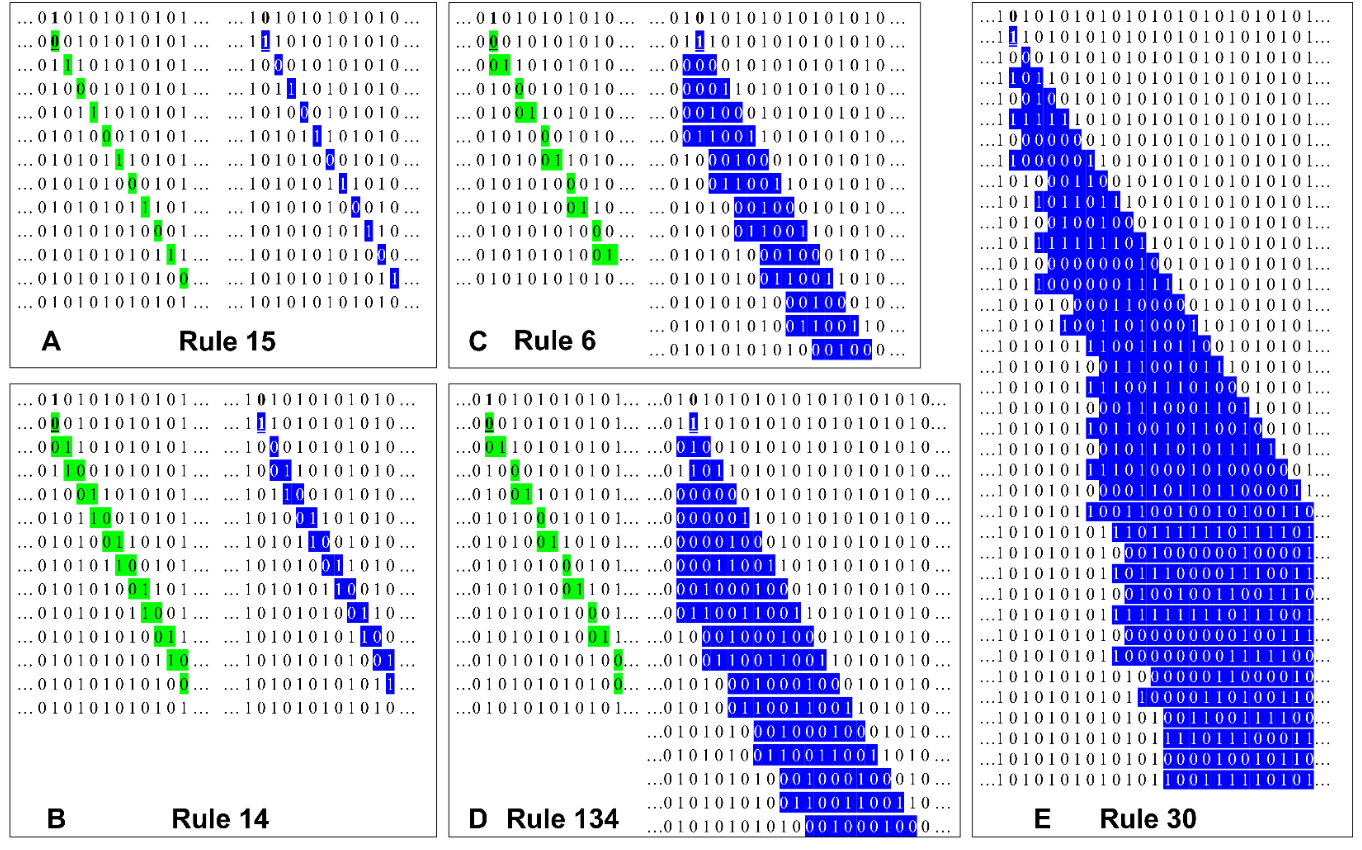}
\caption{Illustrations of propagating perturbation waves in recovering patterns.\ Each row shows a fragment (perturbed) of initially periodic pattern. Rows (top to bottom) show the states of the fragment after each consecutive time step. Perturbed cells are highlighted in \textbf{green/blue} to distinguish between the initial perturbations from state ``1" to state ``0" or from ``0" to ``1" respectively. \textbf{A:} Each perturbed cell recovers in one time step. \textbf{B:} Each perturbed cell recovers in two time steps. \textbf{C: Green case} - perturbed ``0" recovers in one time step and perturbed ``1" - in two. \textbf{Blue case} - perturbed ``0" recovers in five and perturbed ``1" in six time steps. \textbf{D: Green case} - same as on Panel C. \textbf{Blue case} - perturbed ``0" recovers in nine and perturbed ``1" in ten time steps. \textbf{E:} Recovery is slow and completely irregular.}
\label{fig2}
\end{figure}

The perturbation waves propagate from LEFT to RIGHT in the case of rules 6, 14, 15, 30, 134, 135, 143, 158 and 159 and from RIGHT to LEFT for the remaining rules.\ The most important question about the propagating waves of perturbations is at what speed they propagate and, consequently, how long it takes for the perturbation to reach the chain's border where it usually vanishes.\ Our study shows that for rules 15 and 85 (which form a symmetry group, see Table~\ref{table2}), the perturbed cell recovers in one time step and at the same time the perturbation shifts one cell (to the right/left in the case of rule 15/85 respectively).\ Thus, the size of the wave is 1 cell and the speed of the perturbation wave is 1 cell/time-step (see Fig~\ref{fig2}) and, therefore, it takes no more than $n$ time steps (where $n$ is a size of the chain) for the perturbations to reach the chain edge and disappear.

Properties of the perturbation waves forming under rules 14, 143, 84 and 213 (form a symmetry group, see Table~\ref{table2}) are similar to what was observed for rules 15 and 85.\ However, now it takes two time steps for each perturbed cell to recover (see Fig~\ref{fig2}B). The recovery (that is, the contracting border of the perturbed area) as well as the perturbation (that is, the expanding border of the perturbed area) shifts one cell per timestep and the number of perturbed cells (i.e.\ the size of the wave) is 2 cells. When the perturbation reaches the edge of the chain, it doesn't always disappear: for example, in the case of rule 14 the state of the second rightmost cell remains perturbed (at ``0") if the state of rightmost cell in the initial periodic pattern is 0 (the chain in the stationary state has ‘000' on its right border). Similar exceptions occur for the other three rules in the symmetry group (see Table~\ref{table2}). 

The recovery of the periodic pattern under rule 6 is more complicated and takes place in one of two scenarios (Fig~\ref{fig2}C). The first scenario takes place when the originally perturbed cell was in the state ``1" and the perturbation changes its state to ``0".\ In this case, the perturbation expands with the speed of 1 cell/time-step and the recovery takes place in a way that two cells recover simultaneously every second time step so that the size of the wave alternates between 1 and 2 cells. This is because it takes two time steps to recover from ``0" to ``1'' and one time step to recover from ``1'' to ``0". However, if the originally perturbed cell was in state ``0" then the recovery is much slower.\ After the initial transition process, the perturbation wave stabilizes, so that each perturbed “0” recovers in five time steps while each ``1" recovers in six. The speed of the perturbation expansion is still 1 cell per time-step, but its size alternates between 5 and 6 cells. In similarity to the case of rule 15 the perturbation after reaching the edge of the chain, doesn't always disappear but the second leftmost cell remains perturbed (at ``0") if the state of leftmost cell in the initial periodic pattern is 0 (the chain in the stationary state ends with ``000" on the right). Similar scenarios are observed for the three other rules (159, 20 and 215) in the symmetry group (see Table~\ref{table2}). The recovery of the periodic pattern under rule 134 is identical to that under rule 6 when the initially perturbed cell was in state ``1" (Fig~\ref{fig2}D). A difference is observed when the state of the initially perturbed cell changes from ``0" to ``1''.\ After the initial transition process when the perturbation wave stabilizes, the recovery turns to be even slower: each perturbed ``0" recovers in 9 time steps while each ``1" – in 10. The speed of the wave is still 1 cell per time-step but its size alternates between 9 and 10 cells.\ This observation naturally extends to rules 158, 148 and 214 forming the symmetry group with rule 134. 

\begin{center}
\begin{tabular}{|c|c|c|c|} 
\hline
\textbf{Rule} & \textbf{Complement} & \textbf{Mirror image} & \textbf{Mirror complement} \\
\hline
\multicolumn{2}{|c|}{\textbf{LEFT-to-RIGHT}} & \multicolumn{2}{|c|}{\textbf{RIGHT-to-LEFT}} \\
\hline
\multicolumn{2}{|c|}{\textbf{15}} & \multicolumn{2}{|c|}{\textbf{85}} \\
\multicolumn{2}{|c|}{Speed of the wave - 1 cell/TS} & \multicolumn{2}{|c|}{Same as 15.} \\
\multicolumn{2}{|c|}{Size of perturbed area - 1 cell} & \multicolumn{2}{|c|}{} \\
\hline
\textbf{14} & \textbf{143} & \textbf{84} & \textbf{213} \\ 
Speed - 1 & Speed - 1 & Speed - 1 & Speed - 1 \\ 
Size - 2 & Size - 2 & Size - 2 & Size - 2 \\ 
(\ldots000) & (\ldots111) & (000\ldots) & (111\ldots) \\
\hline
\textbf{6} & \textbf{159} & \textbf{20} & \textbf{215} \\ 
$1 \to 0$: Speed - 1 & $0 \to 1$: Speed - 1 & $1 \to 0$: Speed - 1 & $0 \to 1$: Speed - 1 \\ 
Size 1/2 & Size 1/2 & Size 1/2 & Size 1/2 \\
(\ldots000) & (\ldots111) & (000\ldots) & (111\ldots) \\
$0 \to 1$: Speed - 1 & $1 \to 0$: Speed - 1 & $0 \to 1$: Speed - 1 & $1 \to 0$: Speed - 1 \\ 
Size - 5/6 & Size - 5/6 & Size - 5/6 & Size - 5/6 \\
(\ldots000) & (\ldots111) & (000\ldots) & (111\ldots) \\
\hline
\textbf{134} & \textbf{158} & \textbf{148} & \textbf{214} \\ 
$1 \to 0$ (same as 6) & $0 \to 1$ (same as 159) & $1 \to 0$ (same as 20) & $0 \to 1$ (same as 215) \\ 
$0 \to 1$: Speed - 1 & $1 \to 0$: Speed - 1 & $0 \to 1$: Speed - 1 & $1 \to 0$: Speed - 1 \\ 
Size - 9/10 & Size - 9/10 & Size - 9/10 & Size - 9/10 \\
(\ldots000) & (\ldots111) & (000\ldots) & (111\ldots) \\
\hline
\textbf{30} & \textbf{135} & \textbf{86} & \textbf{149} \\ 
Same as 149. & Same as 149. & Same as 149. & Speed of expansion 1. \\
& & & Recovery is slow and irregular.\\
& & &  Size is unlimited. \\
\hline
\end{tabular}
\captionof{table}{Summary on the recovery dynamics for rules allowing formation of propagating perturbation waves.\ The direction, speed and size of the waves formed are given.\ If the properties of perturbation waves depend on the form of original perturbation, they are stated separately: the notation ``$a \to b$" is used to indicate that the originally perturbed cell changed from the state ``$a$" into the state ``$b$".\ The exceptional cases, when the recovery is not complete (the perturbation wave hits the border where the perturbation remains), are described in the extra (last) line.\ For example, the notation ``(\ldots000)" indicates that the recovery is not complete when the perturbed chain ends with three 0s.}
\label{table2}
\end{center}

The recovery of the periodic pattern under rules 30, 135, 86 and 149 (forming a symmetry group) is significantly different from what we have seen so far (see Fig~\ref{fig2}E). The perturbed area under these rules still expands with the speed of 1 cell/time-step (i.e.\ under rule 30, the right border of the disturbed area shifts 1 cell/time-step) but the recovery is much slower: the contracting border (i.e.\ the left border under rule 30) of the perturbed area is slow, is affected by oscillations and its motion is completely irregular. As a result, the perturbed area quickly expands until it reaches the border of the chain, i.e.\ its size is only limited by the size of unrecovered area in the chain (i.e.\ the area to the right of the contracting border under rule 30).\ Although the contracting border moves much more slowly than the expanding one, it can be shown that its speed can't be less than 2 cells per 9 time-steps.\ The properties of propagating perturbation waves under all the rules allowing their formation are summarized in Table~\ref{table2}.

\subsection{Formation of two-periodic patterns from random initial condition}

In the last two sections, we dealt with the rules which allow the recovery of preset periodic patterns perturbed by a rare and random noise.\ In this section, we shall focus on a subset of those rules which not only allow the recovery of perturbed periodic pattern but also the generation of periodic patterns from any initial conditions.\ These are the following six rules: 15, 30, 85, 86, 135 and 149. \par These six rules form two symmetry groups: (15, 85) and (30, 86, 135 and 149) (see Table~\ref{table2}).\ The process of periodic pattern formation is significantly different for these two groups: under rules 15 and 85 periodic patterns are generated considerably faster than under the rules forming the second group.  

\begin{figure}[H]
\centering
\includegraphics[width=0.8\textwidth]{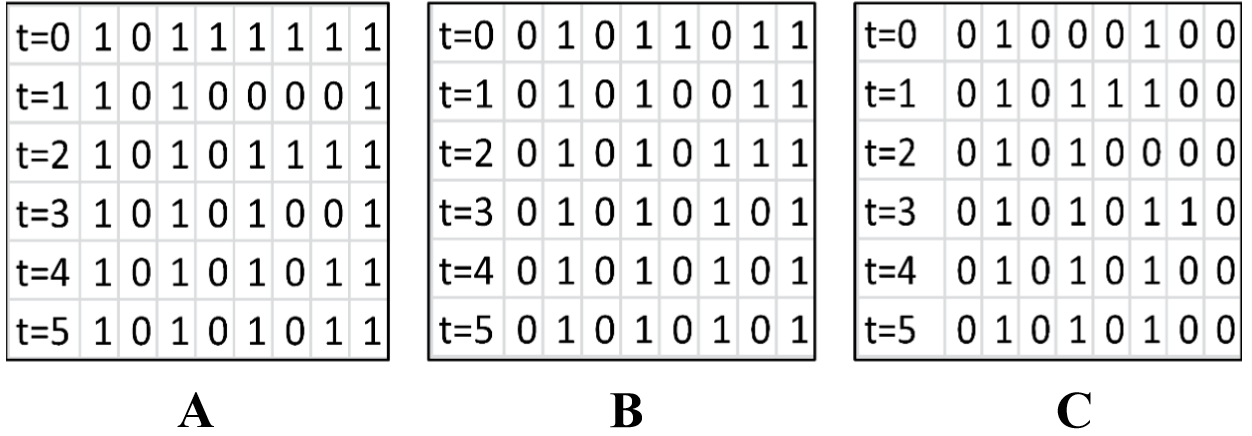}
\caption{Formation of periodic patterns from random initial conditions under rule 15.}
\label{fig3}
\end{figure}   

Three examples of short simulations illustrating the formation of stationary periodic patterns under the rule 15 are shown in Fig~\ref{fig3}. One can see that periodic patterns form in the course of time, first arising at the right border and then expanding to the left with a speed of 1 cell/time-step.\ For rules 15 and 85, one can formulate propositions like the one below: \par

\textit{\textbf{Proposition for rule 15:} Starting from any initial conditions, the state of the chain evolves towards two-periodic pattern (010101\ldots or 101010\ldots) developing from the LEFT to the RIGHT with a speed of at least 1 cell/time-step. The final stationary pattern may end with ``11" or ``00" on its RIGHT edge.} \par

Proof of this proposition is omitted. Proposition for rule 85 differs only in the directionality of the process: under the rule 85 the periodic pattern evolves from right to left.\ The formation of patterns under the other four rules (30, 86, 135 and 149) is considerably different and in a line with the information provided in Table~\ref{table2}.\ For example, in the case of rule 30 the formation of a periodic pattern starts at the right border and develops to the left. The speed at which the periodic pattern expands is irregular and alters from 4 cells per time-step to 2 cells per 9 time-steps.\ Furthermore, depending on the initial conditions, the final pattern may remain non-stationary and exhibit oscillations at the right border of the chain.\ The oscillating spot can contain one element, $(\ldots0100) \leftrightarrow (\dots0110)$, two elements $(\ldots1001) \leftrightarrow (\ldots1111)$ or even three elements $(\ldots11111) \to (\ldots10001) \to (\ldots11011) \to (\ldots10011) \to (\ldots11111)$.\ The periodicity of 11 oscillations is 2 for the first and second cases and 4 for the last case.\ The properties of patterns forming under rules 86, 135 and 149 are similar to those for rule 30.   

\subsection{Three- and more- periodic patterns in ECA}

\begin{figure}[H]
\centering
\includegraphics[width=0.8\textwidth]{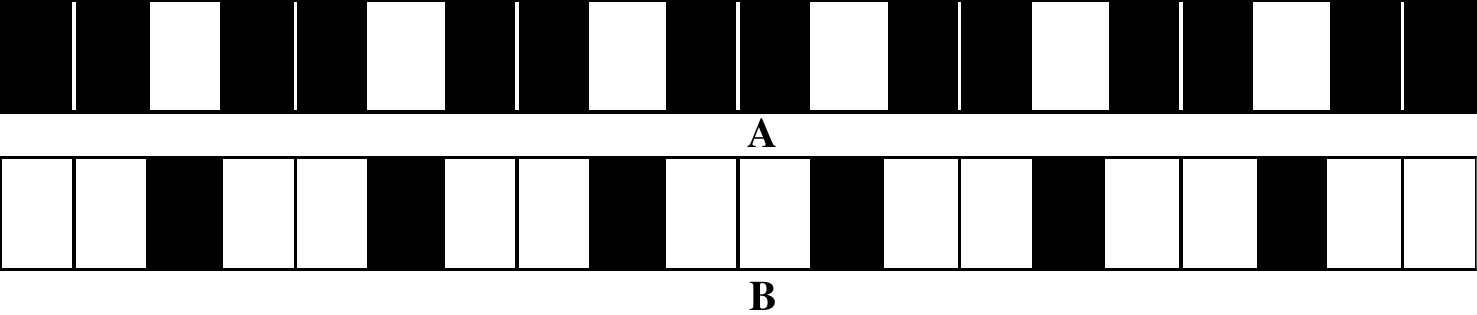}
\caption{Three-periodic patterns in ECA composed of 20 cells. \textbf{A:} Repetitions of 2 black and 1 white cells. \textbf{B:} Repetitions of 2 white and 1 black cells.}
\label{fig4}
\end{figure}

So far, we have focused on rules allowing existence/formation of two-periodic patterns.\ In this section, we address the same questions with respect to three-periodic patterns.\ Two types of three-periodic patterns can be set in ECA: two black cells followed by a white cell or two white cells followed by a black (see Fig~\ref{fig4}).  

Consider the periodic pattern formed by the repetition of one white and two black cells.\ There are 32 rules which do not destroy preset periodic pattern of this type.\ These rules are: 

\begin{center}
4, 5, 12, 13, 36, 37, 44, 45, 68, 69, 76, 77, 100, 101, 108, 109, 132, 133, 140, 141, 164, 165, 172, 173, 196, 197, 204, 205, 228, 229, 236 and 237. 
\end{center}     

All these numbers have binary expressions in the form of $xxx0x10x$ where $x$ is either 0 or 1.\ That is, the pattern consisting of a repetition of 2 blacks and 1 white is conserved under the rules keeping the configurations 2, 3 and 5 unaltered: $(001) \to (001)$, $(010) \to (010)$ and $(100) \to (100)$. 

Similarly, there are 32 rules which conserve a periodic pattern containing one black and two white cells.\ The binary expressions of these rules have the form of $x10x1xxx$.\ That is, the pattern formed by repetitions of 2 whites and 1 black is conserved under the rules keeping the configurations 4, 6 and 7 unaltered: $(011) \to (011)$, $(101) \to (101)$ and $(110) \to (110)$.\ One can notice that there are four rules (76, 77, 204 and 205) which belong to both groups and allow the existence of both types of three-periodic patterns. \par Our next task is to identify rules which allow recovery of three-periodic pattern perturbed by a random and rare noise.\ Our simulations show that there are only two such rules: rule 133, which allows recovery of three-periodic patterns formed by two blacks and one white, and its complement rule 94, which allows recovery of the second type of three-periodic pattern.\ Note that the mirror image of rule 133 is itself and its mirror complement is rule 94.\ Let's consider the recovery of perturbed three-periodic pattern under rule 133 (whose binary representation is 10000101).\ Consider the fragment of periodic pattern \ldots001001001\ldots and assume that the noise strikes it at either 2\textsuperscript{nd} or 3\textsuperscript{rd} or 4\textsuperscript{th} position.\ The perturbed sequence becomes either \ldots011001001\ldots or \ldots000001001\ldots or \ldots001101001\ldots respectively.\ Simple analysis shows that it takes 3, 2 or 3 time steps respectively to recover for each of the cases considered.\ Thus, the recovery under the rule 133 is local.\ All results obtained for rule 133 are naturally extended to rule 94. \par We found no rules which allow the recovery of three-periodic patterns by means of propagating waves.\ Furthermore, there are no rules which allow the formation of threeperiodic patterns from random initial conditions. \par It is easy to show that there are three types of strictly four-periodic patterns which can form in ECA. Indeed, the four-periodic patterns in a generic form can be represented as $\left(a_1a_2a_3a_4\right)^n$ where elements ai can be either 0 or 1 and the chain has $4n$ cells.\ We have 16 possible cases for four-periodicity, namely: 0000, 0001, 0010, 0011, 0100, 0101, 0110, 0111, 1000, 1001, 1010, 1011, 1100, 1101, 1110 and 1111.\ The first and last cases (0000 and 1111) are trivial and can be considered as one-periodic.\ Cases 0101 and 1010 make the classical two-periodic pattern and we have studied them previously.\ All other cases, after re-indexing, fall into one of the three cases: $(0111)^n$, $(0011)^n$ or $(0001)^n$.\ There are 16 rules which can keep the preset 4-periodic pattern of each type.\ The four-periodic pattern of the type $(0111)^n$ is kept by rules in the form $110x1xxx$ (where $x$ is either 0 or 1); patterns of the type $(0011)^n$ - in the form $x1x01x0x$, and patterns of the type $(0001)^n$ - in the form $xxx0x100$.\ None of these rules allow noise resistance or generation of a strictly four-periodic pattern from random initial conditions. \par In similarity to the four-periodic patterns, the patterns of higher periodicity can exist under some rules but they are never noise resistant or can they be generated from random initial conditions.\ The number of rules keeping preset periodic patterns gets smaller with an increase of the periodicity.\ Obviously, the pre-set pattern of any periodicity exists under rule 204: this is an identity rule and doesn't impose any change to any pattern.

\section{Discussion}

The study presented here was motivated by the problem of biological pattern formation governed by local (cell-to-cell) signals.\ As our primary interest was about segmentation in fly embryos, we have focused on the formation of periodic patterns.\ Furthermore, since the segmentation takes place along the embryonic anterio-posterior axis and is essentially a one-dimensional process, we have modelled it using a one-dimensional chain of logical elements represented by ECA.\ In the framework of this model, we have identified all rules allowing the existence or formation of periodic patterns.\ At the same time, our study was mainly focused on properties of ECA and, we believe, that it contributes to an extensive research~\cite{Gravner2012} performed by a large community of scientists in this field. 

In the presented study, we have explored whether formation of morphogenetic patterns in developing tissue can be modelled using ECA.\ The two-state model used in this study represents a chain of locally interacting cells, where cells in state ``1" express some particular gene while in state ``0" - don't.\ Interactions of logical elements in ECA can be considered as representing contact (membrane-to-membrane) interactions between cells in biological tissues.\ The modelled interactions can be seen as regulating the differentiation of cells.\ For example, the two-periodic stationary pattern forming in the model represents a chain of cells where each second cell expresses a certain gene.\ We have identified sets of interaction rules which allow existence, recovery from noise and formation from random initial conditions of various periodic patterns in CA.\ Particularly, we have found that 64 (out of 256) rules allow existence of two-periodic patterns, 33 of them allow recovery of two-periodic patterns from rare and random noise and only six rules (15, 85, 30, 86, 135 and 149) allow formation of two-periodic patterns from any initial conditions. Furthermore, our study of three- and more- periodic patterns has shown that although there are many rules allowing their existence, only two rules allow the recovery of three-periodic patterns from random and rare noise and none allow their recovery from noise or formation from random initial conditions. 

Here we have modelled the formation of periodic patterns in one-dimensional medium of constant size with fixed boundary conditions. Although biological development takes place in three-dimensional objects the formation of simplest (i.e. periodic) patterns can be considered as one-dimensional process. Patterning in a growing tissue implies that the size of the modelled medium is changing over time and modelling the formation of periodic patterns in growing medium is a task for future studies. Also, developing communities of bacteria often form circular chains to model which one should apply periodic boundary conditions. Most of the results obtained in the model with fixed boundary conditions remain unchanged for the model with periodic boundary conditions. Particularly, all results summarised in Table~\ref{table1} remain correct. Rules listed in Table~\ref{table2} allow the recovery of two-periodic patterns provided that the medium is composed of even number of cells. This excludes rules 30, 135, 86 and 149, for which single perturbations transform into chaotic patterns. There are extra interesting results obtained for the periodic boundary conditions.  For example, under rule 142, any perturbation circulates along the looped chain infinitely many times (while in the case of fixed boundary conditions, cells in the right-hand side border do not recover). This result is valid for mirror image and mirror complement of rule 142, namely, for rule 212. 

It is known that the pattern associated with the expression of segmentation genes in the fly embryo is four-periodic.\ As this pattern forms in a few steps and under the influence of maternal, gap and pair-rule genes, one could consider this pattern as preset four-periodic which is allowed by a number of rules.\ However, it is also known that the periodic pattern formed by segmentation genes undergoes certain correction in position~\cite{Jaeger2004} and this wouldn't be allowed by any of transition rules in two-state model.  

\begin{figure}[H]
\centering
\includegraphics[width=1.0\textwidth]{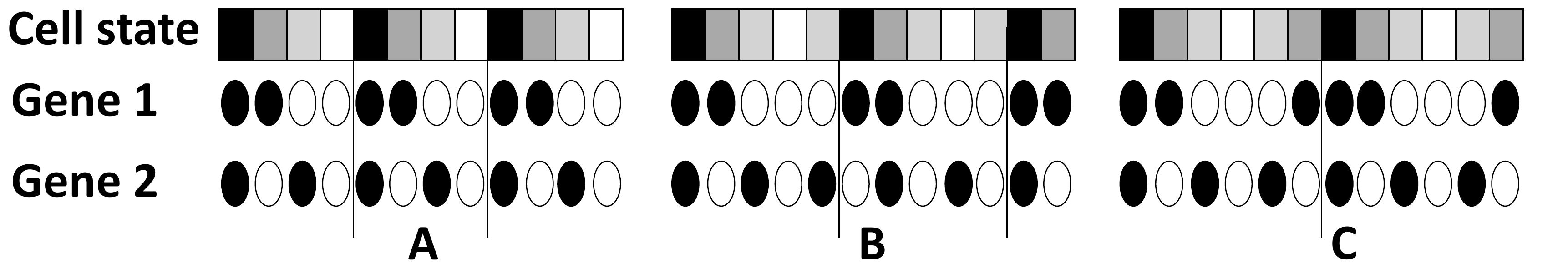}
\caption{Periodic patterns in a four-state model.\ Four- (A), five- (B) and six- (C) periodic patterns with illustration of how each state corresponds to different expressions of two genes.\ Cell state is indicated in black if both genes are expressed; in dark grey - Gene 1 is expressed and Gene 2 is not; in light grey - Gene 1 is not expressed and Gene 2 is expressed; white - both genes are not expressed.}
\label{fig5}    
\end{figure}   

Four-periodicity can easily be obtained if the model allows at least three distinct states for cells.\ While the three-state model doesn't have direct biological implementation, the four-state model seems to have better perspective as it can be viewed as modelling cells whose differentiation is associated with expression of pair of genes.\ The fourperiodic pattern can correspond to the alternation of expressions of two genes as illustrated in Fig~\ref{fig5}A.\ The four-state model is considerably more sophisticated compared with the two-state model and can be used to reproduce the formation four- and even more- periodic patterns (Fig~\ref{fig5}).\ While the number of possible transition rules in the two-state model is relatively small (256),\ this number is extremely large for the four-state model ($4^{4^3}$, that is, every transition rule is a combination of $64=4^3$ elementary rules, while each elementary rule has four forms).\ This number makes the analysis of 4-state models challenging.\ Our preliminary simulations show that the formation of periodic stationary patterns in four-state model is extremely sensitive to the initial conditions, i.e.\ these patterns form only when very special initial conditions are met.\ This may explain the multi- (four-) level of segmentation in the fly embryo.\ The model can account for interactions between segment polarity genes with specific initial conditions set by the above three levels of patterning. \par Mathematical study of the segmentation in fly embryo performed in~\cite{Reka2003} and based on gene network modelling has indicated that all nuclei, during normal development, can be only in four distinct states.\ This points to the direction of further extension of the presented work:\ our extensive analysis of ECA can be considered as a step towards similar analysis of four-state CA.\ Also, we note that more sophisticated CA (two-dimensional and including randomization of transition rules) have recently been used to successfully model the dynamics of skin patterning in growing lizard~\cite{Manukyan2017}. 
\\
\\
\textbf{Acknowledgements.}\ This work was funded by BBSRC grant BB/K002430/1.\ Authors are grateful to Dr Igor Potapov for fruitful discussions.   

\bibliographystyle{unsrt}   
\bibliography{discrete1}

\begin{thebibliography}{10}

\bibitem{Wolpert2002}
Lewis Wolpert, Elliot Meyerowitz, Peter Lawrence, Rosa Beddington, Thomas
  Jessell, and Jim Smith.
\newblock {\em Principles of development}.
\newblock Oxford University Press, USA, 2 edition, 2002.

\bibitem{Murray2003}
James~D Murray.
\newblock {\em Mathematical Biology II: Spatial Models and Biomedical
  Applications}.
\newblock 3rd edition, 2003.

\bibitem{Vasiev2013}
O.~Vasieva, M.~Rasolonjanahary, and B.~Vasiev.
\newblock Mathematical modelling in developmental biology.
\newblock {\em Reproduction}, 145(6):R175--84, 2013.

\bibitem{Hake2003}
Sarah Hake and Fred Wilt.
\newblock {\em Principles of Developmental Biology}.
\newblock W. W. Norton and Company, New York, NY, 2003.

\bibitem{Johnston1992201}
D~St~Johnston and C~Nüsslein-Volhard.
\newblock The origin of pattern and polarity in the drosophila embryo.
\newblock {\em Cell}, 68(2):201—219, January 1992.

\bibitem{Reka2003}
R{\'e}ka Albert and {Hans G.} Othmer.
\newblock The topology of the regulatory interactions predicts the expression
  pattern of the segment polarity genes in drosophila melanogaster.
\newblock {\em Journal of Theoretical Biology}, 223(1):1--18, jul 2003.

\bibitem{Manukyan2017}
Liana Manukyan, Sophie~A Montandon, Anamarija Fofonjka, Stanislav Smirnov, and
  Michel~C Milinkovitch.
\newblock A living mesoscopic cellular automaton made of skin scales.
\newblock {\em Nature}, 544(7649):173—179, April 2017.

\bibitem{Vonneumann1948}
J~von Neumann.
\newblock The general and logical theory of automata.
\newblock 1948.

\bibitem{Vonneumann1966}
John~G. Kemeny.
\newblock Theory of self-reproducing automata. john von neumann. edited by
  arthur w. burks. university of illinois press, urbana, 1966. 408 pp., illus.
  \$ 10.
\newblock {\em Science}, 157(3785):180--180, 1967.

\bibitem{Wainer2010}
Gabriel Wainer, Qi~Liu, Olivier Dalle, and Bernard~P. Zeigler.
\newblock Applying cellular automata and devs methodologies to digital games: A
  survey.
\newblock {\em Simulation \& Gaming}, 41(6):796--823, 2010.

\bibitem{Yang2010}
Xin-She Yang and Y.~Young.
\newblock Cellular automata, pdes, and pattern formation.
\newblock 2010.

\bibitem{Wolfram2002}
Stephen Wolfram.
\newblock {\em A new kind of science}.
\newblock Wolfram Media Inc., 2002.

\bibitem{Wolfram1983}
Stephen Wolfram.
\newblock Statistical mechanics of cellular automata.
\newblock {\em Rev. Mod. Phys.}, 55:601--644, Jul 1983.

\bibitem{Wolfram1986}
S.~Wolfram.
\newblock {\em Theory and Applications of Cellular Automata: Including Selected
  Papers, 1983-1986}.
\newblock Advanced series on complex systems. World Scientific, 1986.

\bibitem{Gravner2012}
Janko Gravner and David Griffeath.
\newblock Robust periodic solutions and evolution from seeds in one-dimensional
  edge cellular automata.
\newblock {\em Theoretical Computer Science}, 466:64 -- 86, 2012.

\bibitem{Jaeger2004}
Johannes Jaeger, Svetlana Surkova, Maxim Blagov, Hilde Janssens, David Kosman,
  Konstantin~N Kozlov, Manu, Ekaterina Myasnikova, Carlos~E Vanario-Alonso,
  Maria Samsonova, David~H Sharp, and John Reinitz.
\newblock Dynamic control of positional information in the early drosophila
  embryo.
\newblock {\em Nature}, 430(6997):368—371, July 2004.

\end{thebibliography}

\newpage

\appendix
\numberwithin{table}{section}

\section{Rule 30} \label{appA}

In this appendix, we analyse the evolution of perturbed periodic pattern under the rule 30 and show that the pattern recovers from perturbation by driving the perturbed area to the right at a speed of at least 2 cells (one strip) per 9 time-steps.\ As a starting point, we note that a single perturbation can result to formation of a large perturbed area and therefore, we will consider a sequence in the form $(01)kxxx\ldots$ (or $1(01)kxxx\ldots$), that is a sequence starting with $k$ stripes followed by an arbitrary com-bination of 0s and 1s.\ We check how many time steps it takes for any such combination to transform to $(01)n+1xxx\ldots$, that is, to form an extra strip.\ We note that the transformation $(01)nxxx\ldots\to(01)n+1x\ldots$ is equivalent to the transformation $1xxx\ldots \to 101x\ldots$ and in the following will refer to the latter case (to reduce notations).\ Below, we split all sequences having the form $1xxx\ldots$ into (ten) groups so that the sequences listed in each following group take one more step for the required transformation. 

\begin{itemize}

\item \textbf{1\textsuperscript{st} group}: sequence is already in the form $101x\ldots$, and thus the transformation took place in zero time steps.  

\item \textbf{2\textsuperscript{nd} group}: sequences resulting in the formation of $101xx\ldots$ in one time step have the form $1100x\ldots$ (i.e.\ $1100x\ldots \to 101xx\ldots$ in one time step). 

\item \textbf{3\textsuperscript{rd} group}: sequences resulting in the formation of $1100xx\ldots$ in one time step (and therefore resulting in the formation of $101xxx\ldots$ in two time steps) have the form $10000x\ldots$. 
\item \textbf{4\textsuperscript{th} group}: sequences resulting in the formation of $10000xx\ldots$ in one time step (and therefore resulting in the formation of $101xxx\ldots$ in three time steps) have the forms $11111xx\ldots$ and $111101x\ldots$.

\item \textbf{5\textsuperscript{th} group}: sequences resulting in the formation of $11111xxx\ldots$ and $111101xx\ldots$ in one time step (and therefore resulting in the formation of $101xxx\ldots$ in four time steps) have the forms $10010xxx\ldots$ and $1001100x\ldots$.

\item \textbf{6\textsuperscript{th} group}: sequences resulting in the formation of $10010xxx\ldots$ and $1001100x\ldots$ in one time step (and therefore resulting in the formation of $101xxx\ldots$ in five time steps) have the forms $11011xxx\ldots$, $110101xx\ldots, 11010000x\ldots, 111000xx\ldots$ and $1110011xx\ldots$.

\item \textbf{7\textsuperscript{th} group}: sequences resulting in the formation of $11011xxx\ldots$ and $110101xx\ldots$ in one time step (and therefore resulting to the formation of $101xxx\ldots$ in six time steps) have the form $10001xxx\ldots$, while no any sequence results to formation of $11010000x\ldots, 111000xx\ldots$ and $1110011xx\ldots$.

\item \textbf{8\textsuperscript{th} group}: sequences resulting in the formation of $10001xxx\ldots$ in one time step (and therefore resulting in the formation of $101xxx\ldots$ in seven time steps) have the forms $11101xxx\ldots$ and $111100xx\ldots$.

\item \textbf{9\textsuperscript{th} group}: sequences resulting in the formation of $11101xxx\ldots$ and $111100xx\ldots$ in one time step (and therefore resulting in the formation of $101xxx\ldots$ in eight time steps) have the forms $1001101xxx\ldots$ and $100111xxx\ldots$.

\item \textbf{10\textsuperscript{th} group}: sequences resulting in the formation of $1001101xxx\ldots$ and $100111xxx\ldots$ in one time step (and therefore resulting in the formation of $101xxx\ldots$ in nine time steps) have the forms $11010001xxx\ldots, 1101001xxx\ldots$ and $111001xxx\ldots$.

\end{itemize}

We note that all sequences having the form $1xxx\ldots$ have appeared in one of the above ten groups and therefore the transformation never takes more than 9 steps.\ A summary of the time required for the transformation $1xxx\ldots \to 101x\ldots$ for all possible sequences is given in the Table~\ref{tableA1}. 

\begin{center}
\begin{tabular}{||l|l||l|l||} 
\hhline{|t:==:t:==:t|}
$10000xxx\ldots$ & 2 time steps & $11010000\ldots$ & 5 time steps \\ 
				 &  			   & $11010001\ldots$ & 9 time steps \\
				 &  			   & $1101001x\ldots$ & 9 time steps \\
				 &  			   & $110101xx\ldots$ & 5 time steps \\
\hhline{||--||--||}
$10001xxx\ldots$ & 6 time steps & $11011xxx\ldots$ & 5 time steps \\
\hhline{||--||--||}
$10010xxx\ldots$ & 5 time steps & $111000xx\ldots$ & 5 time steps \\ 
			  	 &  			   & $111001xx\ldots$ & 9 time steps \\
\hhline{||--||--||}
$1001100x\ldots$ & 4 time steps & $11101xx\ldots$ & 7 time steps \\ 
$1001101x\ldots$ & 8 time steps &                          &                    \\ 
$1001100x\ldots$ & 8 time steps &                          &                    \\ 
\hline
$101xxxxx\ldots$ & 0 time steps & $11110xx\ldots$ & 7 time steps \\ 
                          &                   & $11111xx\ldots$ & 3 time steps \\
\hhline{||--||--||}
$1100xxx\ldots$  & 1 time step  & $11111xxx\ldots$ & 3 time steps \\
\hhline{|b:==:b:==:b|}
\end{tabular}
\captionof{table}{Summary of the number of time steps required for sequences in the form $1xxx\ldots$ to transform to the sequence $101x\ldots$ for rule 30.\ Transforming sequences are indicated in the 1\textsuperscript{st} and 3\textsuperscript{rd} columns, while the number of time steps required for their transformation in the 2\textsuperscript{nd} and 4\textsuperscript{th} columns.}
\label{tableA1}
\end{center}

\newpage

\section{Rule 15} \label{appB}

In this appendix, we analyse the formation of two-periodic patterns under the rule 15 in a chain imposed to the random initial conditions.\ We note that under this rule, periodic patterns form first on the left border and then expand to the right.\ Therefore, we start with the consideration of various initial combinations on the left side of the chain and analyse how the periodic structure forms and expands to the right.\ First, we consider initial conditions with the digit 1 on the left border of the chain.

\begin{itemize}
\item \textbf{Case A:} $101xxx\ldots$ It is easy to see that none of these three digits will change over time.\ The first digit 1 is fixed (fixed boundary).\ The second digit, 0, stays unaltered following the configuration rule $(101) \to (101)$.\ The third digit, 1, also doesn't change no matter what digit is on the forth position following either $(010) \to (0\textbf{1}0)$ or $(011) \to (0\textbf{1}1)$.\ Thus, the given initial condition is consistent with the periodic pattern and conserved. 

\item \textbf{Case B:} $100xxx\ldots$ The second digit, 0, remains unaltered following the configuration: $(100) \to (1\textbf{0}0)$. The third digit, 0, changes to 1 following either $(000) \to (0\textbf{1}0)$ or $(001) \to (0\textbf{1}1)$.\ Thus, we obtain $1\textbf{01}xxx\ldots$ (\textbf{case A}) in one time step:\ $100xxx\ldots \to 1\textbf{01}xxx\ldots$. 

\item \textbf{Case C:} $110xxx\ldots$ The second digit, 1, turns into 0 following $(110)\to(1\textbf{0}0)$.\ The third digit remains unaltered following either $(100)\to(1\textbf{0}0)$ or $(101)\to(1\textbf{0}1)$.\ Thus, in one time step we obtain $100xxx\ldots$ which is \textbf{case B}.\ Thus, it takes two time steps to transfer from \textbf{case C} to (stable) \textbf{case A:} $110xxx\ldots \to 100xxx\ldots \to 1\textbf{01}xxx\ldots$.

\item \textbf{Case D:} $111xxx\ldots$ The second digit, 1, turns into 0 following $(111) \to (1\textbf{0}1)$.\ The third digit, 1, changes to 0 following either $(110) \to (1\textbf{0}0)$ or $(111) \to (1\textbf{0}1)$.\ Thus, we obtain $100xxx\ldots$ (which is \textbf{case B}) in one time step and therefore obtain $101xxx\ldots$ (which is \textbf{case A}) in two time steps: $111xxx\ldots \to 100xxx\ldots \to 1\textbf{01}xxx\ldots$. 

\item \textbf{Cases E-H:} These cases deal with the initial conditions in the form $0xxx\ldots$, i.e.\ with 0 on the left border.\ It is easy to show that the first three digits in the sequence $010xxx\ldots$ do not change over time, which is similar to \textbf{case A} considered above.\ The sequence $011xxx\ldots$ transfers into $010xxx\ldots$ in one time step (similar to \textbf{case B}), while the sequences $000xxx\ldots$ and $001xxx\ldots$ transfer into $0\textbf{10}xxx\ldots$ in two time steps (similarly to \textbf{cases C} and \textbf{D}). \par It is also easy to see that the pre-existing strips on the left side of the chain are conserved, that is, $(10)^nxxx\ldots \to (10)^nxxx\ldots$ and $(01)^nxxx\ldots \to (01)^nxxx\ldots$.\ This observation, in the combination with the above \textbf{cases A-H}, indicates that the periodic pattern forming on the left expands to the right ($(10)^nxxx\ldots \to (10)^{n+1}xxx\ldots$ or $(01)^nxxx\ldots \to (01)^{n+1}xxx\ldots$) and every next strip appears in at least two time steps.\ Thus, the slowest way of formation of periodic structure implies its expansion to the right at a speed of 1 cell per time-step.\ The formation of periodic pattern may remain incomplete if its expansion to the right leads to occurrence of $xxx011$ or $xxx100$ where $xxx$ is a periodic fragment (i.e.\ $xxx=(10)^n$) and therefore stable.\ This follows from the configuration rules $(011) \to (011)$ and $(100) \to (100)$. 

\end{itemize}

\end{document}